\documentclass[slac_one]{revtex4}
\usepackage{graphicx}
\usepackage{fancyhdr}
\begin{document}
%FERMILAB-CONF-08-444-A

%Title of paper
\title{Results from the CDMS 5-Tower Operation}
\author{Jonghee Yoo (for the CDMS collaboration)}
\affiliation{Fermi National Accelerator Laboratory, Batavia, IL 60510, USA}

\begin{abstract}
Astrophysical observations strongly suggest that non-luminous, nonbaryonic components, so called "Dark Matter", may constitute most of the matter in the Universe. The Cryogenic Dark Matter Search (CDMS) experiment is designed to detect Dark Matter interaction events through nuclear recoils from elastic scattering. The detector is capable of reading out both phonon and ionization energy of an interaction in Ge or Si crystals. We present results from the CDMS five-tower detector arrays. The data were collected in the period between October 2006 and July 2007 (with an effective exposure of 121.3 kg-days). No WIMP signal was observed. The results, when combined with previous CDMS Soudan data, set a 90\% confidence level upper bound on the WIMP-nucleon cross section of $4.6\times 10^{-44}$ cm$^2$ at 60 Gev/c$^2$ WIMP mass.
\end{abstract}
\maketitle
\thispagestyle{fancy}

\section{INTRODUCTION} % Section title should be in all capitals.
 There is increasing evidence that a non-luminous, nonbaryonic component, ``Dark Matter'', constitutes most of the matter in the Universe \cite{Cosmol}.  All this evidence, however, is indirect, meaning the hypothesis of unseen Dark Matter can explain anomalies in astrophysical observations. There is no evidence that Dark Matter couples to normal matter other than by gravitational interaction. Its particle characteristics such as composition, spin, mass are all unknown. Any kind of hypothetical neutral and stable non-Standard Model particle can thus be a Dark Matter candidate. Hence, direct observation of Dark Matter will not only answer fundamental questions from astrophysics, but it would also open up a new era of beyond the Standard Model in particle physics. This is one of the most urgent problem that needs to be addressed in the next decades. 

There is no lack of Dark Matter candidates in a theoretical point of view. Among the candidate, the Weakly Interacting Massive Particle (WIMP) is the most interesting Cold Dark Matter candidate for two reasons \cite{WIMP}. First, WIMPs naturally appear in many beyond-the-Standard Model scenarios, such as the Lightest Super-Symmetric Particle (LSP) \cite{LSP} and the Lightest Kaluza-Klein Particle (LKP) \cite{LKP}. Second, the expected range of interaction cross sections with normal matter and WIMP masses are experimentally accessible with existing technology. Therefore, the majority of the direct detection experiments aim to observe WIMPs\cite{Lewin}.

\section{CDMS EXPERIMENT}
The Cryogenic Dark Matter Search (CDMS) experiment is designed to detect WIMP signals through nuclear recoil in the target material by elastic scattering. The CDMS Z(depth)-sensitive Ionization and Phonon detectors measure both the ionization and the phonon signal from the crystal \cite{zips}. The detector is an ultra-pure Ge (250\,g) or Si (100\,g) crystal in a cylindrical shape of 1\,cm thick and 7.6\,cm diameter. A tower consists of vertically stacked six ZIP detector-array. Five towers of detector-arrays, a totaling 30 detectors (19 Ge and 11 Si) have been operating since October 2006. The detectors are cooled down to 50\,mK. An electron-hole pair-inducing interaction in the crystal gives the ionization signal. The electron and hole pairs are separated by an electric field through the crystal (3\,V/cm in Ge and 4\,V/cm in Si). The ionization signals are read out by inner and outer electrodes. The phonon signals are read out by Quasi-particle-assisted Electrothermal-feedback Transition-edge sensors (QETs) on each detector. The Transition-Edge-Sensors (TESs) are voltage biased and a SQUID array reads the current through them. The phonon signals are read out by electro-thermal-feed-back in the TESs. In order to reduce muon-induced neutron backgrounds, the CDMS detectors are installed at the Soudan Underground Laboratory, 780\,m underground, Minnesota, USA. Lead blocks and polyethylene panels are used for passive shielding of background photons and neutrons. Scintillation panels entirely cover the outer detector area for active muon veto. The detector performance is monitored on a real-time basis.

\section{DATA ANALYSIS AND RESULTS}
\begin{figure}[t!]
  \begin{center}
    \includegraphics[height=2.55in]{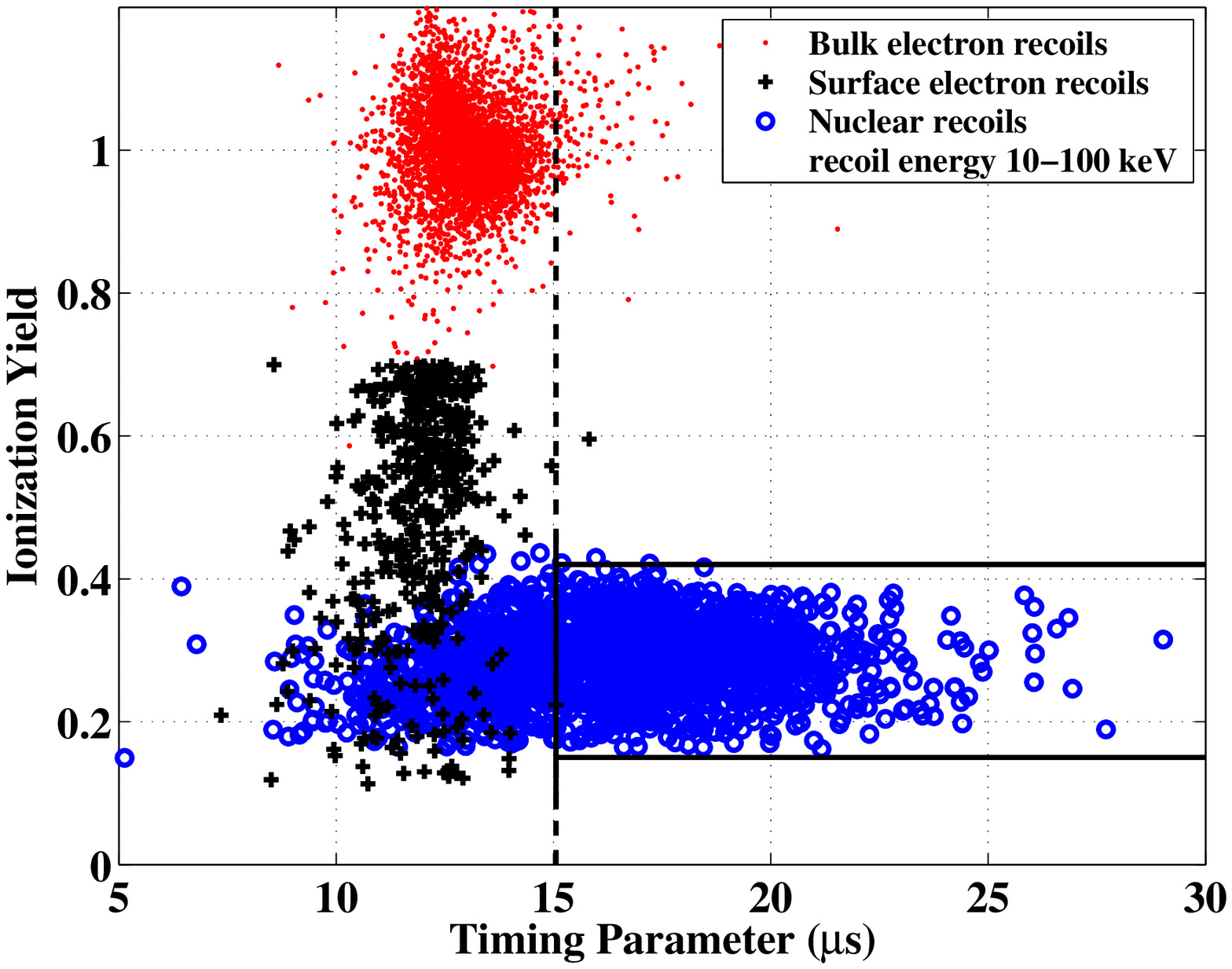} 
    \includegraphics[height=2.50in]{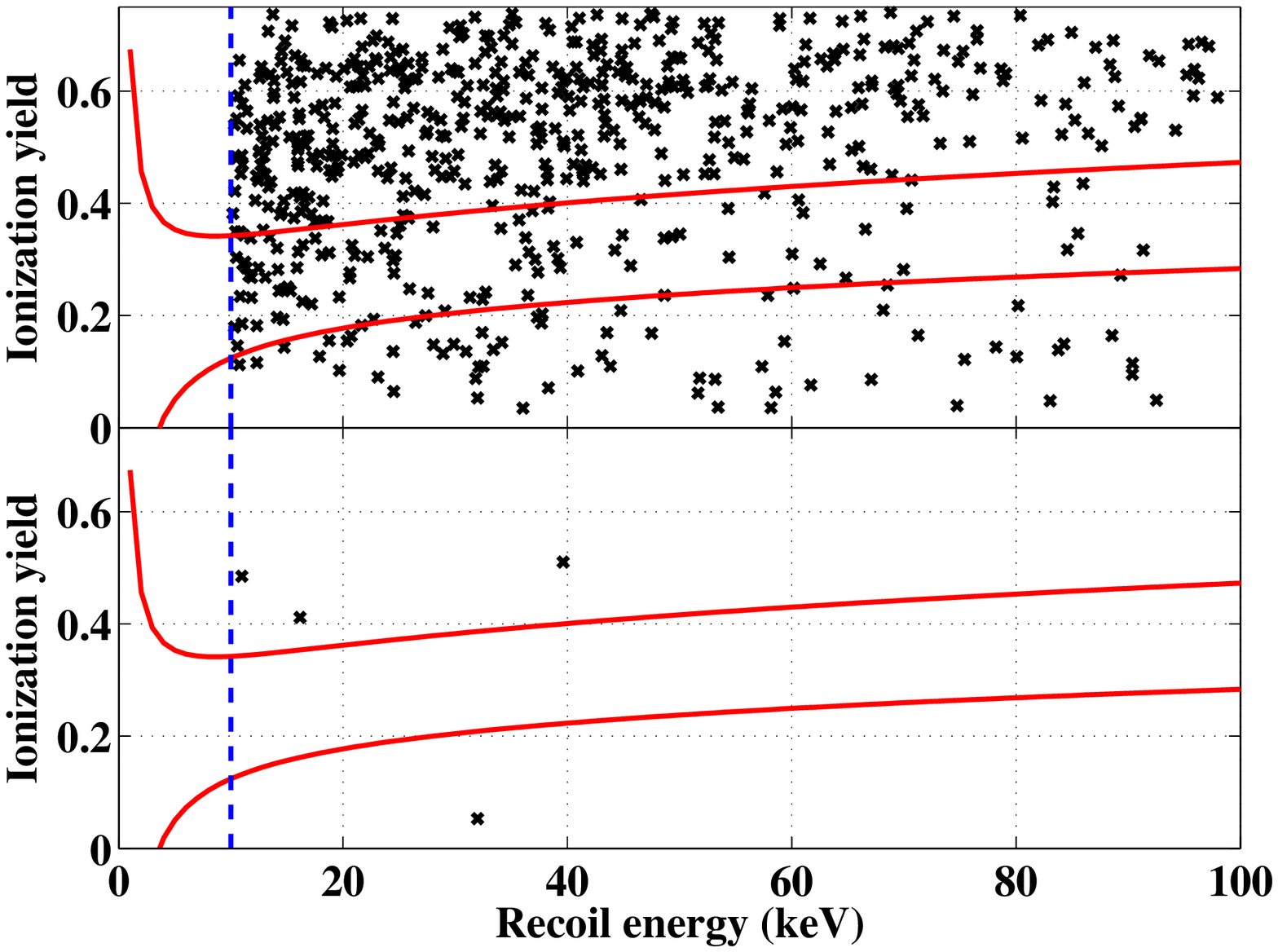}
    \caption{\small  {\bf [Left figure]} Ionization yield versus timing parameter for calibration data in one of the Ge detectors. The yield is normalized to unity for typical bulk-electron recoils\, (dots; from $^{133}$Ba gamma rays). Low-yield $^{133}$Ba events\,(+), attributed to surface electron recoils, have small timing parameter values, allowing discrimination from neutron-induced nuclear recoils from $^{252}$Cf\,($\circ$), which show a wide range of timing parameter values. The vertical dashed line indicates the minimum timing parameter allowed for candidate Dark Matter events in this detector, and the box shows the approximate signal region, which is in fact weakly energy dependent. 
      {\bf [Right figure]} Top: Ionization yield versus recoil energy in all detectors included in this analysis for events passing all cuts except the ionization yield and timing cuts. The signal region between 10 and 100\,keV recoil energies was defined using neutron calibration data and is indicated by the curved lines. Bulk-electron recoils have yield near unity and are above the vertical scale limits. Bottom: Same, but after applying the timing cut. No events are found within the signal region.}
    \label{fig:timing:leakage}
  \end{center}
\end{figure}

This report is based on data collected in the period between October 2006 and July 2007. The two consecutive cryogenic runs during this period are named r123 and r124. In this analysis we consider the 15 good Ge detectors (for WIMP search) for r123 and 7 good Ge detectors for r124. Gamma ($^{133}$Ba) and neutron ($^{252}$Cf) sources were used for energy calibrations and WIMP search efficiency studies. Quality cuts were carefully defined to identify data sets with significant deviations from normal condition. Energy and position dependence of the timing of the phonon signals were corrected using a look-up table based on the electron recoil calibrations. 

In order to be a Dark Matter candidate event, the deposited energy should be 4$\sigma$ above mean noise level. As the WIMPs will not interact more than once in our detector, we require only single scatter events and no significant activity in the surrounding scintillator veto shield. The events should belong to the 2$\sigma$ region of the nuclear recoil band in ionization yield. The phonon rise time and pulse delay from the ionization signal is used to discriminate surface electron induced events. The left plot in Figure~\ref{fig:timing:leakage} shows an example of the distribution of the ionization yield versus timing parameter in the calibration data. Photons can be clearly discriminated from nuclear recoil events by comparing its ionization yield distribution. However, some electrons drop down to the nuclear recoil area (black dots in the figure). The fast timing characteristics of those electron events can be used to make a pure neutron signal area, hence a nuclear recoil events signal area (black box region in the figure).

\begin{figure}[t!]
  \begin{center}
\includegraphics[width=4in]{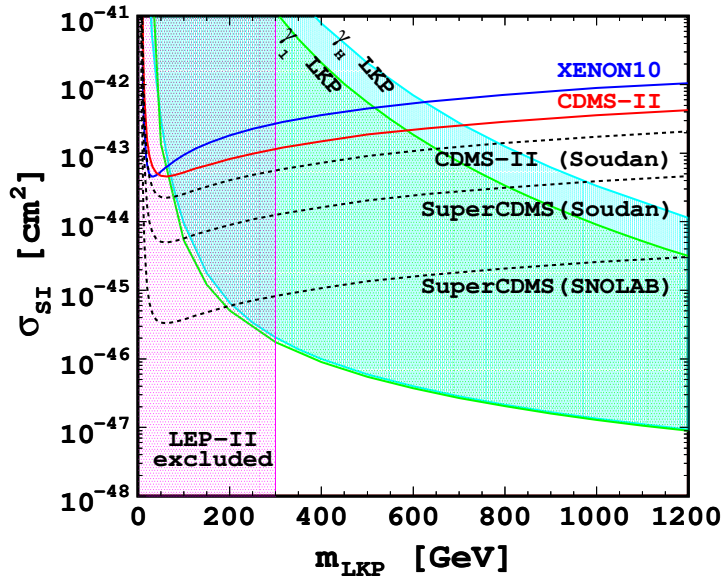}
\caption{\label{fig:limit}
  \small The new upper limit from the CDMS experiment is shown in red solid curve. The previous best bound from XENON-10 is drawn in blue solid curve. The CDMS-II goal (Soudan), SuperCDMS 15\,kg (Soudan) and 100\,kg (SNOLAB) phases are drawn in black dashed curves. The allowed regions for Kaluza-Klein photons in 5D ($\gamma_1$, green shaded) and 6D ($\gamma_H$, blue shaded) is shown for various mass splittings \cite{KKDM:2008}. The bound from LEP-II is shaded in magenta.}
\end{center}	
\end{figure}

A blind analysis was carried out. The events in the WIMP signal region were masked before defining the final cuts. The WIMP selection criteria and background estimation in the signal region were finalized before looking at the WIMP search region. The expected background due to surface electron interactions is estimated to be $0.6\pm0.5$ events.
Neutrons induced by cosmic ray muons or radioactive processes were estimated using Monte Carlo simulations. Less than 0.1 cosmogenic neutron background is expected in the signal region. Neutron background induced by nuclear decay, such as from U and Th chain is estimated to be less than 0.1\, event. 

Most of the quality cuts have little affect on the WIMP detection efficiencies. The ionization-based fiducial volume ($\sim$30\% cut out) and phonon-timing cuts ($\sim$30\% cut out) are the dominant factor of the WIMP detection efficiencies. The effective exposure before the cuts is 397.8\,kg-days, and the net exposure after applying all the cuts is 121.3\,kg-days (averaged over recoil energies 10--100\,keV, weighted for a WIMP mass of 60\,GeV/c$^2$).

On February 4, 2008, the WIMP search signal region was unmasked. No events were observed in the WIMP search signal region while the total expected background in the signal region was less than 0.6 event. Therefore, the results are consistent with null observation of WIMPs, or none of the non-Standard Model particles were observed within the detector sensitivity. The right-side plots in Figure~\ref{fig:timing:leakage} shows the low-yield events observed in all detectors used in this analysis. The upper panel shows the ionization yield distribution versus energy for single-scatter events passing all selection cuts except the timing cut. The four events passing the timing cut shown in the lower panel are outside the 2$\sigma$ nuclear recoil region. 

This null observation result sets an upper bound on the WIMP-nucleon cross section and mass of the WIMPs parameter space. Combining all Soudan data, the upper limit is $4.6 \times 10^{-44}$\,cm$^2$ at 90\% C.L. for a 60\,GeV/c$^2$ WIMP. Figure~\ref{fig:limit} shows the upper bound from CDMS (red solid curve) and XENON-10 (blue solid curve) \cite{Angle:2007uj}. Since the expected WIMP parameter space from Super-Symmetric Dark Matter candidates (mostly neutralinos) has been shown many times in our previous publications \cite{cdms:previous}, here we show the Kaluza-Klein Dark Matter (KKDM) parameter space for minimal UED models \cite{KKDM:2008}. In this minimal model there are only two free parameters, the mass coupling between Lightest Kaluza-Klein Particles (LKP) and KK-quarks ($\Delta$m, here 0.01  $\le \Delta$m $\le$  0.5) and the mass of Higgs (m$_H$, here m$_H$=120\,GeV). The dotted curves indicate projected detector sensitivities of CDMS-II Soudan goal, SuperCDMS 15\,kg at Soudan and SuperCDMS 100\,kg at SNOLAB. The SuperCDMS project will start to probe a substential part of the Kaluza-Klein Dark Matter parameter space.

\section{SUMMARY}
We presented results from the CDMS five-tower detector arrays with a net exposure of 121.3\,kg-days after cuts. No WIMP signal was observed. The result set an upper bound for WIMP-nucleon cross section at 4.6$\times10^{-44}$\,cm$^2$ when combined with previous CDMS Soudan data in 90\%C.L. for a WIMP of mass 60\,GeV/c$^2$. This is the best WIMP search sensitivity above a WIMP mass of 42\,GeV. CDMS is the only experiment that has achieved zero background in the WIMP signal region, which means CDMS has demonstrated the highest power of Dark Matter discovery potential. Continuous running of the current detector (CDMS-II at Soudan) and future Super detectors (SuperCDMS at Soudan and SNOLAB) will provide substantial improvement to the WIMP search sensitivity.

\begin{acknowledgments}
The CDMS collaboration gratefully acknowledges Patrizia Meunier, Daniel Callahan, Pat Castle, Dave Hale, Susanne Kyre, Bruce Lambin and Wayne Johnson for their contributions.  This work is supported in part by the National Science Foundation (Grant Nos.\ AST-9978911, PHY-0542066, PHY-0503729, PHY-0503629,  PHY-0503641, PHY-0504224 and PHY-0705052), by the Department of Energy (Contracts DE-AC03-76SF00098, DE-FG02-91ER40688, DE-FG03-90ER40569, and DE-FG03-91ER40618), by the Swiss National Foundation (SNF Grant No. 20-118119), and by NSERC Canada (Grant SAPIN 341314-07).
\end{acknowledgments}

\end{document}